\input harvmac
\input epsf
\def\figin#1{#1}
\def\ifig#1#2#3{\xdef#1{fig.~\the\figno}
\goodbreak\midinsert\figin{\centerline{#3}}%
\smallskip\centerline{\vbox{\baselineskip12pt
\advance\hsize by -1truein\noindent\footnotefont{\bf Fig.~\the\figno:} #2}}
\endinsert\global\advance\figno by1}

\def\vecx{\vec{x}}
\def\vecv{\vec{v}}

\def\vece{\vec{E}}
\def\vecb{\vec{B}}

\def\vecv{\vec v}

\def\rst{R_*}

\lref\callan{C.G. Callan and J.M. Maldacena, ``Brane Dynamics from the
Born-Infeld Action,'' Nucl. Phys. B513 (1998) 198, hep-th/9708187.}
\lref\gibbons{G.W. Gibbons, ``Born-Infeld Particles and Dirichlet P-Branes,''
Nucl. Phys. B514 (1998) 603, hep-th/9709027.}
\lref\howe{P.S. Howe, N.D. Lambert, P.C. West, ``The Self-Dual String Soliton,''
Nucl. Phys. B515 (1998) 203, hep-th/9709014. }
\lref\lee{S. Lee, A. Peet and L. Thorlacius, ``Brane-Waves and Strings,'' 
Nucl. Phys. B514 (1998) 161, hep-th/9710097.}
\lref\rey{S.-J. Rey and J. Yee, ``Macroscopic Strings as Heavy Quarks in Large
N Gauge Theory and Anti-de Sitter Supergravity,'' hep-th/9803001.}
\lref\bak{D. Bak, J. Lee and H. Min, ``Dynamics of BPS States in the
Dirac-Born-Infeld Theory,'' 
Phys. Rev. D59 (1999) 045011, hep-th/9806149.}
\lref\savvidy{K. G. Savvidy and G. K. Savvidy, ``Neumann Boundary Conditions from 
Born-Infeld Dynamics,'' hep-th/9902023.}
\lref\schwarz{J.H. Schwarz, ``The Power of M Theory,'' 
Phys. Lett. B367 (1996) 97, hep-th/9510086. }
\lref\jackson{J.D. Jackson, ``Classical Electrodynamics, 2nd Edition,'' John Wiley \& Sons,
Inc. (1975), equation 9.18.} 
\lref\leigh{R.G. Leigh, ``Dirac-Born-Infeld Action from Dirichlet Sigma Model'', 
Mod. Phys. Lett. A4 (1989) 2767.}
\lref\gkmt{A. Gomberoff, D. Kastor, D. Marolf and J. Traschen, ``Fully Localized Intersecting
Branes - The Plot Thickens," hep-th/9905094.}




\Title{\vbox{\baselineskip12pt
\hbox{hep-th/9906237}}}
{\vbox{\centerline{\titlerm Dynamics of the DBI Spike Soliton}
 }}
{\baselineskip=12pt
\centerline{David Kastor\foot{kastor@phast.umass.edu} and 
Jennie Traschen\foot{traschen@phast.umass.edu}}
\bigskip
\centerline{\sl Department of Physics and Astronomy}
\centerline{\sl University of Massachusetts}
\centerline{\sl Amherst, MA 01003-4525}}
\bigskip
\medskip
\centerline{\bf Abstract}
\bigskip
We compare oscillations of a fundamental string ending on a D$3$-brane in two
different settings: (1) a test-string radially threading the horizon of an extremal black
D$3$-brane and (2) the spike soliton of the DBI effective action for a D3-brane. 
Previous work has shown that overall transverse modes of the test-string appear 
as $l=0$ modes of the transverse scalar fields of the DBI system. 
We identify DBI world-volume degrees of freedom that have dynamics matching those of the
test-string relative transverse modes. We show that there is a map, resembling
$T$-duality, between relative and overall transverse modes for the test-string that interchanges
Neumann and Dirichlet boundary conditions and implies equality of the absorption coefficients for
both modes. We give general solutions to the overall and relative transverse parts of the
DBI coupled gauge and scalar system and calculate absorption coefficients for the higher angular
momentum modes in the low frequency limit.  We find that there is a nonzero
amplitude for $l>0$ modes to travel out to infinity along the spike, demonstrating that the spike
remains effectively $3+1$-dimensional. 
\Date{June, 1999}

\newsec{Introduction}
The DBI world-volume field theory of a D$p$-brane has BPS soliton solutions, first studied
in \callan\gibbons\howe, that carry both electric and scalar charge. Because 
the world-volume scalar fields record the spacetime position of the brane, the diverging
scalar at the soliton core means that the surface of the brane has a spike reaching off to
infinity, as illustrated in the figure below.
Both the soliton's shape and its electric charge suggest that it
represents a fundamental string ending on the D$p$-brane. This interpretation is supported
\callan\ by a matching of the divergent self-energy of the soliton with the mass of a
semi-infinite string. In addition it has been shown
\gkmt\ that coupling of the spike soliton to bulk supergravity fields yields the 
Neveu-Schwarz $B_{\mu\nu}$ field appropriate to a fundamental string at linear order.
\ifig\fone{The basic spike soliton. The world-volume $r$-coordinate vanishes 
at the end of the spike, while the spacetime $R$ coordinate goes to infinity.}
{\epsfysize=1.75in \epsfbox{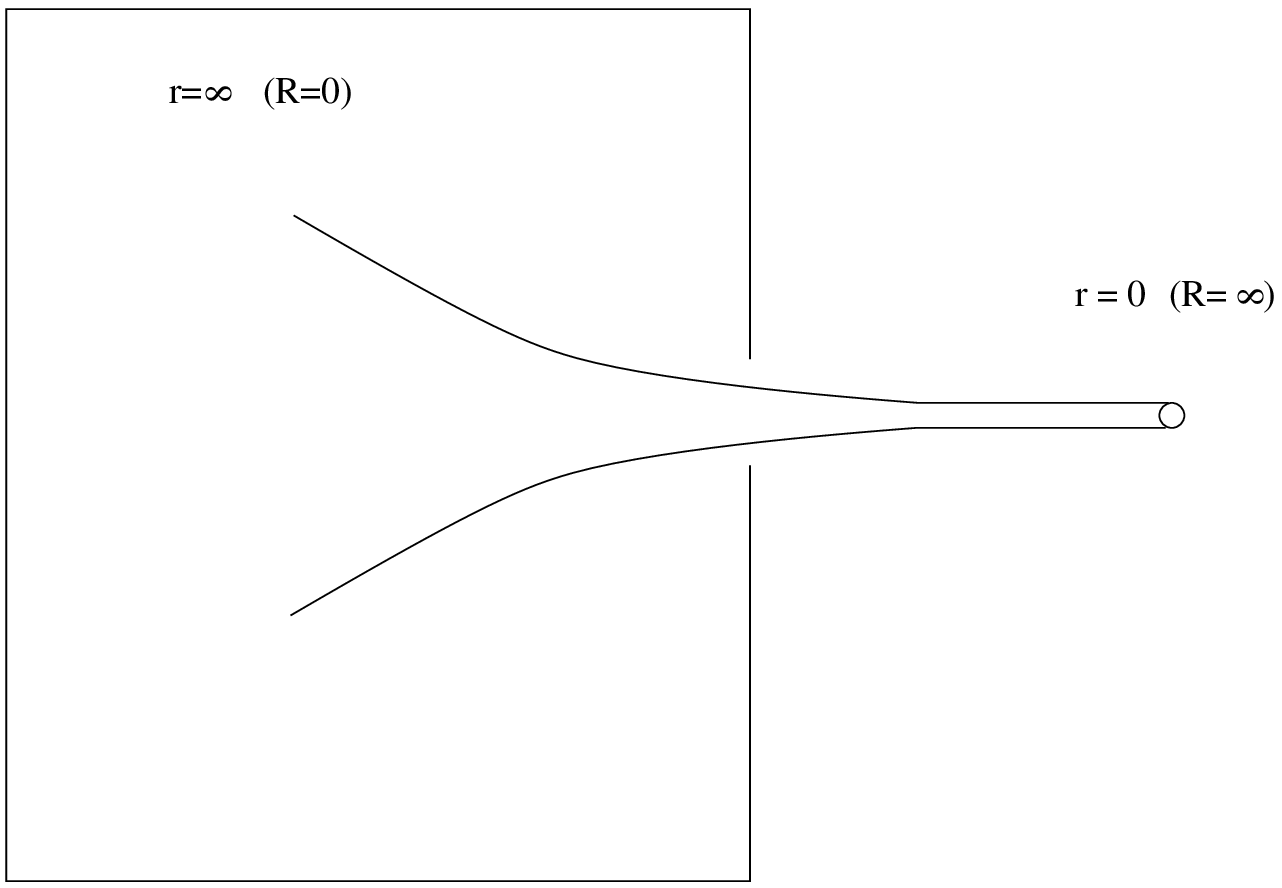}}\vskip -8pt

One can probe this correspondence further by studying the dynamics of the DBI spike
soliton \callan\lee\rey\bak\savvidy . In \lee\ it was shown that for $p=3,4$ the equation
governing oscillations of the spike in directions transverse to the D$p$-brane world-volume is
identical to that for certain oscillations of a test-string threading the horizon of a D$p$-brane
spacetime. 
In addition to these ``overall transverse''
modes, the test-string also has ``relative transverse'' oscillations in directions
transverse to the string but tangent to the D$p$-brane world-volume. Our focus in this
paper will be on finding an identification of these relative transverse modes with degrees of
freedom of the DBI spike soliton for the case of the D$3$-brane. This task is more
complicated than in the overall transverse case, 
because the relevant perturbations of the spike soliton involve the coupled gauge
and scalar field dynamics of the $3+1$-dimensional DBI system in a nontrivial way, which has no
a priori analogue in the $1+1$-dimensional test-string system.

Carrying out the identification 
requires a  detailed understanding of both the test-string and the coupled 
world-volume gauge and
scalar field dynamics.   As we work through this material we 
highlight a number of interesting aspects of both the test-string and DBI spike soliton systems.
For example, in section (2) we note the existence for the test-string of a map, bearing a
striking resemblence to $T$-duality, that takes overall transverse modes $X_\perp$into relative
transverse modes $Y_{||}$ and vice-versa. The map implies equality of the 
absorption coefficients
for the two types of modes. Another curious feature of the test-string system is an
$R\rightarrow 1/R$ symmetry that exchanges the near horizon and
asymptotically flat regions \savvidy.  In the DBI system, we show that such a symmetry continues
to hold for all angular momentum modes and exchanges the flat part of the brane with the region
down the spike. 

We give low frequency approximations
to the absorption coefficients for all modes of both systems. For the DBI system this includes
modes carrying nonzero angular momentum.  We note that, although 
suppressed by an angular momentum barrier, the transmission of these modes down the spike is
nonzero. Therefore, even though the spike gradually takes on
characteristics of a fundamental string, it remains effectively $3$-dimensional all the way out to
infinity.  The presence of the higher angular momentum modes illustrates a basic difference
between the DBI spike soliton and spacetime test-string systems.

Our main result is an understanding of how the relative transverse modes are manifest in the
DBI system.  Physically, an oscillation of the spike soliton parallel to the brane looks like a
linearly oscillating charge. Therefore, we should expect to see dipole radiation in both the
electromagnetic and scalar fields \savvidy.  We find that this is indeed the case and further that
the moduli scalars describing the center of mass position of the spike soliton satisfy the same
equation as the test-string relative transverse modes.

\newsec{Fluctuating Test String in a D$p$-brane Background}
We start by deriving the equations of motion for fluctuations of a test-string 
in a D$p$-brane background.  The string metric for the D$p$-brane is given by
\eqn\metric{\eqalign{
ds^2&=H^{-1/2}(-dt^2+\dots +dx_p^2)+H^{1/2}(dR^2 +R^2d\Omega _{8-p}^2)\equiv
g_{\alpha\beta}dx^\alpha  dx^\beta.\cr
H&=1+\left({\mu\over R}\right)^{7-p} ,\qquad R^2=x_{p+1}^2+\dots +x_9^2.\cr}}
The Nambu-Goto area
action for the test string is simply $S =\int d^2\sigma \sqrt{-\det\, G_{AB}}$,
where $G_{AB}=\partial_A
x^\alpha\partial_B x^\beta g_{\alpha\beta}$ is the induced metric on the
worldsheet. Note that this string does not couple to the RR gauge field of the
D$p$-brane. A static string stretching radially outward in the $x_9$ direction from the D$p$-brane
horizon at $R=0$ solves the  equations of motion and we study perturbations around it.
Choosing static gauge $\sigma^0=t$ and $\sigma^1=x^9$ (implying $\sigma^1=R$ for the radial
string), to second order in small fluctuations we have 
\eqn\fluctuations{
-det\, G_{AB}\simeq +1-\left\{
(\partial_tx^k)^2-H^{-1}(\partial_Rx^k)^2\right\} -
\left\{ H(\partial_tx^\mu)^2 -(\partial_Rx^\mu)^2\right\},}
where $k,=1,..,p$ and $\mu=p+1,...,8$. Denoting the overall
transverse modes $x^\mu$ by $X_\perp$ and the relative transverse modes by
$Y_{||}$, the equations of motion for these modes are then given by
\eqn\tottrans{
\partial_R^2X_\perp-H\partial_t^2X_\perp =0,  }
\eqn\reltrans{
\partial_R({1\over H}\partial_R Y_{||})-\partial_t^2 Y_{||} =0. } 
Below we will need the energy densities $E_\perp, E_{||}$ and energy
fluxes $F_\perp, F_{||}$ for these wave equations which satisfy the
conservation law $\partial_t E +\partial_R F=0$. These are given by
%
%
\eqn\ttflux{
E_{\perp}= {1\over 2}\left(H\dot X_\perp^2+(\partial_RX_\perp)^2\right),
\qquad F_{\perp} = - \dot X_\perp\partial_RX_\perp} 
\eqn\rtflux{E_{||}= {1\over 2}\left( \dot Y_{||}^2+{1\over H}(\partial_RY_{||})^2\right),
\qquad F_{||}= - {1\over H}\dot Y_{||}\partial_RY_{||}. }

\subsec{An Analogue of $T$-Duality }
By definition, oscillations of the end point of a fundamental string on a D$p$-brane
satisfy Neumann boundary conditions for directions along the brane and Dirichlet boundary 
conditions for perpendicular components. The two types of oscillations are interchanged
by the action of $T$-duality, which also changes the dimension of the brane.

A priori, the overall and relative transverse modes of the test string
$X_\perp$ and $Y_{||}$ are geometrically distinct, probing different components of the
background spacetime metric. Indeed, as will see below, when equations \tottrans\ and \reltrans\
are put in standard scattering form, the potentials are qualitatively different. Surprisingly
however, there exists a simple map relating the two types of modes that bears a striking
resemblance to $T$-duality.
Specifically, making the substitutions
\eqn\map{ \partial _R Y_{||}=
H(R) \partial _t
X_\perp  ,\qquad \partial _t Y_{||}= \partial _R X_\perp  }
brings equation \tottrans\ into the form of equation \reltrans. This implies, for example, that 
if $X_\perp$ is a solution of \tottrans\ with time dependence $e^{-i\omega t}$, then 
$Y_{||}=i/\omega\partial_R X_\perp$ solves \reltrans.
 Note that the map \map\
interchanges Dirichlet and Neumann boundary conditions.
Also note that, although $T$-duality changes the dimension of the background brane, because of
the required translation invariance of the lower dimensional brane, the metric function $H(R)$
remains unchanged.

The relation \map\ between overall and relative transverse modes implies that the absorption
coefficients for the two types of modes are identical. If $X_\perp$ and $Y_{||}$ are related by
\map, then the corresponding energy fluxes $F_\perp$ and $F_{||}$ defined above are equal.
The dimensionless absorption coefficient is the ratio of the flux
absorbed into the horizon to the flux incident from infinity
\eqn\abscoeff{ 
\sigma = { F_{hor}\over F_{\infty}^{in} } }
Hence the absorption coefficients $\sigma_\perp$ and $\sigma_{||}$ are equal. 
Note that this is
an exact statement about the governing equations \tottrans\ and \reltrans  .
However, one must keep in mind that the derivation of these equations
from the area action assumes $|\partial_R X|\sim \omega A \ll 1$, where $A$ is the
amplitude of the wave.

\subsec{Absorption Coefficients for $X_\perp$ and $Y_{||}$}
We now specialize to the case of a test-string ending on a D$3$-brane and compute
scattering coefficients $\sigma_\perp=\sigma_{||}\equiv\sigma$ for overall and relative transverse 
waves to be absorbed into, or reflected back from, the horizon \callan. In order to bring
equations \tottrans\ and \reltrans\ into standard scattering form, we change to a tortoise
coordinate $R_*$ defined by $dR_*=\sqrt{H(R)}dR$ and also rescale the mode wavefunctions 
according to 
$ \psi_\perp =H^{+1/4}X_\perp$ and $\chi_{||} = H^{-1/4} Y_{||}$. 
For Fourier modes with time dependence $e^{\pm i\omega t}$, we then have
\eqn\tttort{\partial^2 _{\rst}\psi_\perp +(\omega ^2 -V_\perp (\rst ) )\psi_\perp =0 }
\eqn\rttort{\partial^2 _{\rst}\chi_{||} +(\omega ^2 -V_{||} (\rst ) )\chi_{||} =0 }
where the overall and relative transverse potentials are given by
\eqn\potentials{\eqalign{
V_{\perp} & =H^{-1/4}\partial ^2 _{R_*} (H^{1/4} ) = {5 \mu^4\over R^6}H^{-3}\cr
V_{||} &= H^{+1/4}\partial ^2 _{R_*} (H^{-1/4} ) = \left({-5\mu^4\over R^6}+
{2\mu^8\over R^{10}}\right)H^{-3}\cr
}}
The tortoise coordinate $R_*$ has the asymptotic behavior that $R_*\sim R$ for $R\rightarrow
\infty$, while $R_*\sim -\mu^2/R$ for $R\rightarrow 0$.
 %
The overall transverse potential $V_{\perp}$ can be seen to 
fall off rapidly, like $\rst ^{-6}$, for both $\rst\rightarrow \pm \infty $.
Solutions to the overall transverse equation 
are then well approximated by plane waves 
$\psi_\perp \sim \exp (\pm i\omega\rst )$
in both asymptotic regions. Therefore,
at long wavelengths, $V_{\perp}$ can be approximated by a delta-function potential.
Performing the matching gives the dimensionless absorption coefficient \abscoeff\ \callan
\eqn\abs{
\sigma  = 4(\mu\omega)^2 }
We have demonstrated by means of the `T-Duality' discussed above that this is also the absorption
coefficient for the relative transverse waves.

The near horizon limits of $X_\perp$ and $Y_{||}$
are suggestive of $3+1$ dimensional physics.
Near the horizon the tortoise coordinate $\rst
\rightarrow -\infty$ and we have
\eqn\horeqs{
(\partial ^2 _{\rst} +\omega ^2 )\psi_\perp  \simeq 0,\qquad
(\partial ^2 _{\rst} -{2\over \rst ^2} +\omega ^2 )\chi_{||} \simeq 0.}
The mode wavefunctions in this limit are then 
\eqn\horsols{X_\perp \simeq {1\over \rst} e^{-i\omega \rst},\qquad
Y_{||} \simeq \rst  e^{-i\omega \rst}\left(1-{i\over \omega R_*}\right)}
As the overall transverse wave $X_\perp$ approaches the horizon, it has the characteristic
$1/R_*$ decay of the spherically symmetric mode of a scalar field carrying energy off to
infinity in a $3+1$ dimensional space.
As we will discuss below, this matches the behavior
of the lowest angular momentum mode for overall transverse oscillations of the $3+1$ DBI spike
soliton. 
We also note that, the equation for $Y_{||}$, if regarded as the equation for
a radial function in $3+1$ dimensions, has a dipole nature.
It contains the $l(l+1)/\rst ^2$ term with $l=1$. 
The divergrent asymptotic behavior of the relative transverse waves $Y_{||}$ in \horsols\ is
confusing, but note that the energy flux computed from \rtflux\ remains finite. 
This behavior turns out to be appropriate so that 
$Y_{||}$ may be realized in the DBI spike system as an $l=1$ oscillatory mode
radiating energy to infinity along the D$3$-brane.

\subsec{A Curious $R\rightarrow 1/R$ Symmetry}
We complete this section by noting an interesting symmetry of the
overall transverse wave equation \tottrans\ \savvidy. One can check that if $X_\perp (R)$ solves
\tottrans  , then so does  %
\eqn\dualsol{
\tilde{X}_\perp(R) = R\, X_\perp({\mu^2 \over R} )}
In terms of the tortoise coordinate, this property arises from 
the $R_*\rightarrow -R_*$ reflection symmetry of the scattering potential $V_\perp (\rst )$. 
The mapping \dualsol\ relates the behavior of oscillations of the test string in the AdS near 
horizon region to behavior in the asymptotically flat region. We will return to this
symmetry below when analyzing the DBI modes, where \tottrans\ is generalized to
contain all angular momentum modes, but keeps this symmetry.

\newsec{Fluctuations of the DBI Spike Soliton}
The DBI action for the world-volume degrees of freedom of
a D$3$-brane in a flat background and corresponding equations of motion are given in static gauge by
\eqn\biaction{
S=\int d^4x\sqrt{-det\left(\eta_{\mu\nu}+F_{\mu\nu}\right)},\qquad
\left({1\over \eta-F^2}\right)_\lambda^{\ \nu}\partial_\nu F_\mu^{\ \lambda}=0,}
where the Greek indices $\mu,\nu$ run from $0,\dots,9$. For the D$3$-brane we divide these into
two sets, directions $\alpha,\beta = 0,1,2,3\,$ tangent to the brane and directions
$a,b=4,\dots,9\,$ transverse to the brane. We then have
\eqn\fields{
F_{\alpha\beta}=\partial_\alpha A_\beta-\partial_\beta A_\alpha, \qquad
F_{\alpha b}=-\partial_\alpha\phi_b,\qquad
F_{ab}=0,}
where $A_\alpha$ is the world-volume gauge field and $\phi_a$ are the overall
transverse scalars.
%
%

As shown in \callan\gibbons , the DBI theory has BPS soliton solutions
that represent a collection of spikes projecting outward from the brane. Taking the
spikes to point in the $\phi_9$ direction, the solitonic gauge and scalar fields are
related via $F_\alpha^{\ 0}=\pm\partial_\alpha\phi_9$, and the equations of motion reduce to
$\nabla^2\phi_9=0$. For a single spike at the $3+1$ dimensional origin, we have
$\phi_9=\pm q/r$.

The spike soliton represents a fundamental string ending at an electric charge on
the D$3$-brane. Strictly speaking the string meets the D$3$-brane only at $r=0$,
$\phi_9=\infty$ \leigh . The spike itself has properties which interpolate
between those of the D$3$-brane proper and those of the string. For example, as
we will see below, the spike remains effectively $3+1$ dimensional even as it
narrows near the origin. On the other hand, when coupled to the bulk spacetime
fields, the spike acts as a source for the components of the NS anti-symmetric
tensor field appropriate to a fundamental string \gkmt.

%
%
%
%

Here we will 
catalogue all the fluctuating modes of the DBI spike soliton and
identify the particular modes which correspond to the oscillations
of the test string described in section 2.
 Note that in the DBI coordinates above $r\rightarrow 0$ is going
down the spike to what corresponds to the asymptotically flat region in the
spacetime ($R\rightarrow \infty $).
Let
$F_\mu{}^\nu=\bar F_\mu{}^\nu+\delta F_\mu{}^\nu$, where $\bar F_\mu{}^\nu$ is
the unperturbed spike solution. The perturbed equations of motion
\lee,
\eqn\perturbed{
\bar B_\lambda{}^\nu\partial_\nu(\delta F_\mu{}^\lambda)
+\left(\bar B\left[\bar F\delta F+\delta F\bar F\right]\bar
B\right)_\lambda{}^\nu\partial_\nu\bar F_\mu{}^\lambda =0, \qquad
\bar{B}_\lambda^{\ \nu}=\left({1\over\eta-\bar F^2}\right)_\lambda^{\ \nu}}
yield a system of four equations corresponding to different values for the free index
$\mu=0,k,a,9$, where $k=1,2,3$ and $a=4,\dots,8$. The overall transverse modes $\phi_a$ decouple
from the other fields and satisfy the equation
\eqn\dbitt{
\nabla^2\phi_a-\tilde H(r)\partial _0^2\phi_a =0,}
%
where $\tilde H(r)=1+q^2/r^4$.
The remaining fields give a coupled Maxwell-scalar 
field system, coupled through the background
spike soliton $\phi_9=\pm q/r$
\eqn\lots{\eqalign{
\nabla\cdot\vec E\mp (\tilde H(r)-1)\partial_0^2\psi +
\vec\nabla \tilde H(r)\cdot (\vec E\mp \vec\nabla\psi) &=0,\cr
\vec\nabla\times\vec B- \tilde H(r)\partial_0\vec E
\pm(\tilde H(r)-1)\partial_0\vec\nabla\psi&=0,\cr
\nabla^2\psi - \tilde H(r)\partial_0^2\psi \pm
\vec\nabla \tilde H(r)\cdot (\vec E\mp \vec\nabla\psi)&=0,\cr
}}
%
where $E_k=\delta F_{0k}$, $\delta F_{kl}=-\epsilon_{klm}B^m$,
$\delta F_{kA}=-\partial_k\phi_A$, and $\delta F_{k9}=-\partial_k\psi$.
Note that for $\tilde H(r)=1$ ($q=0$), these equations reduce to the decoupled Maxwell and
transverse scalar equations.

In order to untangle equations \lots\ for the coupled gauge and scalar field 
system, it is useful to define
the combination $\vec v=\vec E\mp\vec\nabla\psi$. Fourier transform in time,
or alternatively consider only an overall time dependence $e^{-i\omega t}$.
  In terms of
$\vec v$, the equations of motion and the Bianchi identities may be rewritten
as the set (see also \bak)
\eqna\bigset
%
$$\eqalignno{
\nabla^2\vec v +\omega ^2 \tilde H(r)\vec v &=0,&\bigset a\cr
\vec\nabla \cdot\vec v &=\pm\omega ^2 \psi,&\bigset b\cr
\nabla^2\psi +\omega ^2 \tilde H(r)\psi &=\mp\vec v\cdot\vec\nabla\tilde H(r) &\bigset c\cr
i\omega \vec B &= \vec\nabla \times\vec v &\bigset d\cr
\vec\nabla\times\vec B & =-i\omega\left( \vec\nabla \psi + \tilde H(r) \vec v
\right)&\bigset e \cr
\vec\nabla\cdot\vec B &=0.&\bigset f\cr}$$
%
Note that the wave equation for each cartesian component $v^k$ is the same
as the equation \dbitt\ for the transverse scalars $\phi_A$. 
Given a solution of \bigset {a}\ for $\vec v$, solutions for the remaining fields
may be determined in the following simple way. 
Let $\vec v$ be a solution to \bigset {a}, then
$\psi$ is given by the right hand side of the Gauss' law constraint \bigset {b}. 
The dynamical equation \bigset {c}\ for $\psi$ is then satisfied as a consequence of \bigset {a}.
Given $\vec v$, the world-volume magnetic field $\vecb$ is determined by
the dynamical equation  \bigset {d}. Equation \bigset {f}\ is then identically satisfied, while
equation \bigset {e}\ follows from the equations above.
Finally, the world-volume electric field is found from via the
definition $\vece =\vecv \pm \vec\nabla\psi$.

The full set of independent solutions to the system \bigset {}\  can then be enumerated
as follows. Each cartesian component of $\vec v$ can be expanded as
$v^k =\sum A^k _{lm}Y_{lm} (\Omega )P_l (r)$, where the functions $P_l(r)$ satisfy a radial
equation which follows from \bigset {a}. The scalar  $\psi$ and the magnetic field
$\vec B$ are then determined as described above in terms of $P_l(r)$ and are indexed by angular
momentum mode $l$. As we will see below, the multipole composition of the full solution 
for a given value of $l$ is actually of a mixed nature.

Both the overall transverse equation \dbitt\ and the relative transverse system \bigset {}\ have
a conserved energy and energy flux vector defined via
${d{\cal E}\over dt} + \vec\nabla\cdot \vec{\cal F} =0 $.
For the overall transverse modes $\phi_a$ we have
\eqn\dbittflux{
{\cal E}_\perp= {1\over 2}\left(\tilde H(r)\dot\phi^2 +\vec\nabla\phi\cdot\vec\nabla\phi\right),
\qquad {\cal F}_\perp=-\dot\phi\vec\nabla\phi ,}
while the relative transverse equations give
\eqn\dbiflux{\eqalign{
{\cal E}_{||}&= {1\over 2}\left(\dot\psi^2 +\vec B^2 + \tilde H(r)\vec v^2\right )
+(\vec\nabla\psi)^2+\tilde H(r)\vec v\cdot\vec\nabla\psi,\cr
\vec{\cal F}_{||}&= \vec{\cal F}_{\psi\psi} +\vec{\cal F}_{\psi v} +\vec{\cal F}_{EB},\cr
\vec{\cal F}_{\psi\psi}&=-\dot\psi\vec\nabla\psi,\qquad 
\vec{\cal F}_{\psi v}=-(\tilde H(r)-1)\dot\psi\vec v,\qquad
\vec{\cal F}_{EB}=\vec E\times\vec B.\cr }}
%

%
%


\subsec{Overall Transverse Fluctuations}
In order to write down the most general solution to the overall transverse wave equation \dbitt,
we decompose the scalars $\phi$ in terms of spherical harmonics,
$\phi=\sum A_{lm}Y_{lm}(\theta,\varphi)P_l(r)$, where $A_{lm}$ are constant coefficients. The
radial equation following from \dbitt\ is then given by 
\eqn\radial{
{1\over r^2}\partial_r(r^2\partial_r P_l)+\left(\omega^2 \tilde H(r)-{l(l+1)\over
r^2}\right)P_l=0,} %
which as $r\rightarrow\infty$ reduces to the usual Helmholtz radial equation. 
If we work instead in terms of the radial coordinate $R=q/r$, then this becomes
\eqn\newradial{
\partial_R^2 P_l+\left (\omega^2 \tilde  H(R)-{l(l+1)\over R^2}\right)P_l=0,\qquad
R={q\over r}.}
For $l=0$, as noted in \lee, this is the same as the equation \tottrans\ governing overall
transverse oscillations of a test-string in a D$3$-brane background, provided that the parameter
$q^2$ that specifies the charge of the spike soliton is identified with the parameter $\mu^4$ from
the D$3$-brane metric. The quantity $\mu^4$ is proportional to the ADM mass per unit volume of the
D$3$-brane spacetime \metric , which is in turn proportional to the D$3$-brane tension $T_3$. The
DBI charge $q$ is proportional the the tension of the attached fundamental string $T_1$. The
equality $q^2=\mu^4$ then requires $T_3 \propto T_1^2$, a result discussed in \schwarz\
that follows from duality arguments involving alternate dimensional reductions of
M-branes. 

We are also interested in the $l>0$ modes of the transverse scalars. 
While the $l=0$ modes of the overall transverse DBI scalars can be indentified 
with overall transverse modes of the test-string, the higher $l$ modes 
present in the DBI system have no obvious analogue amongst excitations of the test-string. 
From equation \radial\ one might think that the higher $l$ modes are suppressed as they
propagate down the spike towards $r=0$. However, a closer analysis shows that, although there is
the usual angular momentum suppression, $l>0$  modes can in fact propagate out to infinity along
the spike. 

Two additional forms of the radial equation are useful for this analysis.  
In terms of the rescaled  radial function $F_l =rP_l$ equation \radial\ becomes %
\eqn\anotherradial{
\partial_r^2 F_l+\left (\omega^2 H(r)-{l(l+1)\over r^2}\right)F_l=0, \qquad 
F_l =rP_l.}
Finally, making use of the D$3$-brane spacetime tortoise coordinate
$\rst$ defined via $d\rst =\sqrt{H(R)}dR$ and rescaling the radial function according to
$P_l =H^{-1/4}\phi _l$ gives
\eqn\tortoise{\partial_{\rst}^2\phi _l +\left(\omega^2-{l(l+1)\over 
H(R)R^2}-V_\perp\right)\phi _l =0,}
where $R$ is regarded as a function of $R_*$ and the potential $V_\perp$ is given in 
equation \tttort  . 
Equation \tortoise, like its $l=0$ case \tttort, is invariant under the reflection $\rst
\rightarrow -\rst$, which interchanges the flat part of the brane near $r=\infty$ with the
region far down the spike near $r=0$.
Both the scattering
potential $V_\perp$ and the angular momentum barrier in \tortoise\ are 
symmetric about the mouth of the spike.  An $l$-mode with a given initial amplitude may be
started at either end, and the subsequent propagation is the same in
either case. In appendix A below, we show that the dimensionless absorption coefficient for 
modes of arbitrary angular momentum $l$ is given in the low energy limit by 
\eqn\absorb{\sigma _l\equiv
{|c_l|^2\over |a_l |^2 }\simeq {4(\omega \sqrt{q})^{4l+2}\over L^2,}}
where $L$ is a numerical constant given in the appendix. We see that although the absorption of
$l>0$ modes is indeed supressed by higher powers of the frequency $\omega$, there is no absolute 
barrier to transmission down the spike. The spike remains functionally $3+1$ dimensional
all the way out to infinity.

We are left with two questions.  What is the
analogue, if any, of the $l\neq 0$ DBI modes in the test string picture? 
And, does the DBI spike soliton actually include a 1+1 dimensional string,
or just the attachment point of the string? One possibility is
that to include the string, one must add the string action to 
the DBI 3-brane action, and find solutions of the combined system. The string would then be
attached at the end of the spike.
Conversely, the test string in the D$3$-brane spacetime approximation
does not include the smooth transition mouth region, which then misses
the higher $l$ modes.

Finally, these DBI modes also display the symmetry that for the test-string relates the physics in
the  near horizon anti-deSitter region to that in the asymptotically flat region.  In the DBI
system, the relation is between the region down the spike and the flat region of the brane.
Note that \anotherradial\  and \newradial\  are the same equation, so that
there is a kind of $r\rightarrow q/r$  symmetry in the
transverse scalar  wave equation. Precisely, if $F_l (r)$ is a solution
to \anotherradial  , then $P_l (r) =r F_l (q/r) $ is also a solution. This
generalizes the symmetry of the spacetime totally transverse modes \dualsol  to
all $l$.

\subsec{Relative Transverse Modes}
There is no direct match among the DBI world-volume fields for the test-string relative
transverse degrees of freedom $Y_{||}$. However, the physical picture is clear \savvidy. The
string ends in an electric charge on the D$3$-brane world-volume, and relative transverse
oscillations of the string result in oscillations of the end point. Since the oscillating
end point carries both electric and scalar charge, we expect to get  both scalar and
electromagnetic radiation.

In the overall transverse case, only the $l=0$ mode of the world-volume scalars corresponded to
the test-string degrees of freedom, and this will also be the case for the relative transverse
modes. As  discussed above, a solution to the relative transverse system of equations 
\bigset {}\ can be specified in terms of a solution of \bigset {a}\ for $\vec v$.
Solutions to the system specified by the $l=0$ modes of $\vec v$, we will see, match the relative
transverse modes of the test-string. We will also see that these modes arise via oscillations of
the moduli of the spike soliton corresponding to its center of mass position.

To illustrate more generally the behavior of the Maxwell-scalar system \bigset {}, we will give
three examples, having $l=0,1,2$ and absorption coefficients that go
like $(\omega\sqrt{q})^{4l+2}$ as $\omega \sqrt{q} \rightarrow 0$. These describe
the analogue of the spacetime relative transverse oscillations, a
dilation of the charge source, and an oscillating electric dipole respectively.

There are three independent $l=0$ modes corresponding to the different
components of $\vec v$ (see also the discussion in \savvidy).  Taking, for example, a wave
polarized in the $z$-direction, we have %
\eqn\voo{v^x =v^y =0,\qquad v^z =P_0 (r).}
Following the steps described above, the other world-volume fields are then given by
\eqn\others{\eqalign{ 
\psi &={1\over \omega^2}P_0 '(r) \cos\theta ,\cr
E^{\hat\theta} &=  -\left(P_0+{1\over \omega^2 r}P_0^\prime\right)\sin\theta,\qquad
E^{\hat r}=E^{\hat\phi}=0,\cr
B^{\hat\phi}&= {i\over \omega}P_0^\prime\sin\theta,\qquad
B^{\hat r}=B^{\hat\theta}=0,\cr}}
where hatted indices indicate components in an orthonormal frame.
We can now compute the flux of radiation $\vec{\cal F}_{||}$ for this solution using the
expressions in \dbiflux . The two contributions $\vec{\cal F}_{\psi\psi}$ and $\vec{\cal F}_{\psi
v}$ each individually diverge as $r$ approaches zero, down the spike. However, their sum is
finite. We find
\eqn\dbifluxzero{\eqalign{
{\cal F}^r_{\psi\psi}+{\cal F}^r_{\psi v} &=
 -{i\over 2\omega} \cos^2 \theta  (P_0^* P_0^\prime -P_0^{* \prime} P_0 ) \cr
{\cal F}^r_{EB} &=
 -{i\over 2\omega} \sin^2 \theta (P_0 ^* P_0^{\prime} -P_0^{*\prime} P_0 ) \cr
{\cal F}^r_{||} &= -{i\over 2\omega}(P_0^* P_0^\prime -P_0^{* \prime} P_0 )\cr
}}
which shows that while radiation in the individual fields exhibits a dipole pattern,
the total radiated energy flux is isotropic.
We also see explicitly that the total radiated flux is identical to that in
the totally transverse case, as expected because 
the solution for the various fields is given in terms of the single function $P$.
This further implies that the absorption coefficient for the $l=0$ relative
transverse DBI excitation is equal to the absorption
coefficient for the test string relative transverse mode, using
the identity of the two totally transverse systems, and the equality $\sigma_\perp=\sigma_{||}$
for the test string absorbtion coefficients.
In fact we have now established on of our main results, the equality of all the absorption
coefficients for the $l=0$ modes of the DBI system with the modes of the test-string
\eqn\equalabs{\sigma _{DBI||}=\sigma _{DBI\perp}=\sigma _{ST||}
=\sigma _{ST\perp}\approx 4(\omega \sqrt{q})^2, }
where the last relation follows from equation \absorb\ and equating $q^2=\mu^4$ as discussed above.

The dynamical degrees of freedom
of the test string and DBI relative transverse modes appear quite different.
For the test string, there are simply the three scalars $Y^k_{||}$, while in the DBI system the
dynamics are described by the coupled $\vece$, $\vecb$ and $\psi$ fields. 
To see how the degrees of freedom
$Y^k_{||}$ arise in the DBI system \savvidy, consider perturbations of the spike soliton
in which the moduli of the spike vary,
\eqn\phimoduli{
\phi _9 ={q\over |\vecx - \Delta\vecx  (t, x^i ) |}.} 
The scalar field $\psi$ is then given by 
\eqn\vary{
\psi\equiv  \delta \phi _9 ={q \vecx \cdot \Delta \vecx\over |\vecx |^3 }.}
We note here that in fact the spike position moduli
$\Delta x^i$ satisfy the same equation \reltrans\ as the relative transverse modes of the
test-string. For example, choosing $\Delta x =\Delta y =0$ yields $\psi =
{q\Delta z cos\theta \over r^2}$ which in turn implies that $v^k$ is of the form assumed in \voo .
Therefore, $\Delta z$ is given by the function $\Delta z ={r^2\over \omega^2 q}P_0 '(r)$. Two
further differentiations then yield the result that $\Delta z$ indeed satisfies equation \reltrans.
Note finally that $\vecv$ and $\vecb$ can be found using $P_0 =-{q\over r^2}H(r)
\Delta z'$, so that the full set of fields is determined by the the behavior of the spike
modulus $\Delta z$.

\subsec{Higher $l$ Modes}
We close this section by presenting two higher $l$ solutions to
the relative transverse DBI system \bigset {}. 
The first is the
spherically symmetric mode of $\vecv$, which is actually the $l=1$ mode of the relative 
transverse system.
Take
$v^k={x^k \over r}P_1 (r)$
which is equivalent to $v^r =P_1 ,v^{\theta} =v^{\phi} =0$. The remaining fields are then
determined to be 
\eqn\vonerest{\eqalign{ 
\psi &={1\over \omega^2}\left(P_1^{\prime} +{2\over r}P_1\right),\qquad
E^r =-{q^2 \over r^4}P_1,\cr E^{\theta} & =E^{\phi} =0,\qquad B^k =0.\cr}}
We see that this mode is a spherically symmetric oscillation of the charged
region. There is therefore no electromagnetic radiation, however there is radiation in the scalar
field $\psi$ with a total flux having the same form  as \dbifluxzero\
\eqn\dbifluxone{ F^r_{tot} =
 -{i\over 2\omega} (P_1 ^* P_1^{\prime} -P_1^{*\prime} P_1 ) }
Using \dbifluxone\ and the expression for the absorption coefficients in 
\absorb\ we have for this mode $\sigma _1 ={4\over 9} (\omega  \sqrt{q})^6$.

Our final example is very similiar to the oscillating dipole in
standard E\&M (see e.g, \jackson). Take $\vecv$ to be a generalization
of a dipole\foot{Recall that this is all in fourier space, so there
are also implicit oscillating factors}
\eqn\vtwo{v^r = 2 cos\theta P_2 (r),\qquad v^{\theta}= {1\over r}P_2(r),\qquad v^{\phi} =0}
This is equivalent to taking the cartesian components $v^k$ proportional to the radial 
function $P_2(r)$ times linear combinations of the spherical harmonics $Y_{20}, Y_{2,\pm1}$.
The scalar field $\psi$ then has a dipole form, there is dipole radiation in
the electromagnetic fields, and the radial component of $\vec E$ is asymptotically
of the same form as in \jackson :
\eqn\vtworest{\eqalign{
\psi & = {2\over \omega^2} \cos\theta \left( P_2^\prime +{3\over r}P_2 \right) \cr
E^{\hat\theta} &=  \sin\theta[P_2-{2\over \omega^2r}(P_2^\prime +{3\over r}P_2)]\cr
B^{\hat\phi}&= -i{\sin\theta\over\omega}(P_2^\prime+{3\over r}P_2)\cr
E^{\hat r}&= {2 cos\theta\over \omega^2} \left({1\over r}P_2^\prime +{3\over r^2}P_2-{\omega^2
q^2\over r^4}P_2\right  ) \cr   }}
In the limit $r\rightarrow\infty$, $E^{\hat r}$ then has the usual dipole form
\eqn\lastequation{
E^{\hat r}\simeq {2\cos\theta\over r^2}e^{i\omega r}.}
%

\bigskip
\noindent
{\bf Acknowledgements:} We thank Vijay Balasubramanian for collaboration on the initial
stages of this work, Amanda Peet for helpful conversations and the Institute for Theoretical
Physics for its hospitality while this work was carried out. This work was supported in part by
NSF grant PHY98-01875 and at ITP by NSF grant PHY94-07194.

\appendix{A}{Scattering Coefficients}
In this appendix we derive
the low energy absorption coefficients for all modes of the DBI system.
Recall that the radial equation \newradial\ governs both overall transverse
oscillations and the oscillations of each component $v^k$ in the relative
transverse DBI systems (as well as the overall transverse test-string modes). 
The radial equation \newradial\  is put in standard scattering form in \tortoise\
by  using the tortoise coordinate $d\rst =\sqrt{H(R)} dR$ and the rescaled wavefunction $\phi
=H^{1/4}P$. Note that this implies that the region
down the spike is $\rst,R\rightarrow \infty$, 
while the flat part of the D$3$-brane is
$\rst\rightarrow -\infty$, $R\rightarrow 0$ and that the angular momentum
term in \tortoise\  falls off like $l(l+1)/\rst ^2 $ as $\rst\rightarrow
\pm\infty$.   The potential
$V_\perp$ falls off like $1/\rst^6$ at either end, implying that solutions to
\tortoise\ are well approximated in terms of Bessel functions  at
the two ends.  We can then write %
\eqn\limits{\eqalign{
\rst\rightarrow\infty;&\qquad \phi_l(\rst)\simeq 
\omega\rst \left( a_l h^{(2)}_l(\omega \rst)
+ s_l h^{(1)}_l(\omega \rst) \right) \equiv\phi_{spike},\cr
\rst \rightarrow -\infty;&\qquad\phi(\rst )\simeq 
\omega\rst c_l h^{(2)}_l(\omega
\rst )\equiv\phi_{out}.\cr}}

The strategy for calculating the absorption coefficients for the different $l$ modes
is to expand  $P_{l,spike}$ and $P_{l,out}$ for small $|\rst|$ and patch them
together using the  $\omega=0$ solution valid in a middle region. Recall that
$\psi=H^{-1/4}\phi$.  Using the asymptotic forms of the Hankel functions and the 
toroise coordinate we then get the expansions 
\eqn\expansions{\eqalign{
P_{l,spike} &\simeq{(a_l+s_l)\over (2l+1)!!}(\omega R)^{l+1}+i 
{(a_l-s_l)(2l-1)!!\over (\omega R)^l}\cr
P_{l,out}& \simeq -\omega\sqrt{q}\left\{{1\over (2l+1)!!}\left(-{\omega q\over
R}\right)^l+ i(2l-1)!!\left(-{R\over \omega q}\right)^{l+1}\right\}c_l,\cr}}
where $(2l+1)!!\equiv (2l+1)(2l-1)\cdots (5)(3)(1)$.
For $\omega=0$ we have
\eqn\middle{\partial_R^2P_{l,middle}-{l(l+1)\over R^2}P_{l,middle}=0,}
which is solved by
\eqn\midfunction{\psi_{l,middle}=\alpha_lR^{l+1}+\beta_l{1\over R^l}.}
Matching coefficients  across the three regions yields the relations
\eqn\matching{\eqalign{
a_l+s_l & = i(-1)^l(\omega \sqrt{q})^{-(2l+1)}L c_l \cr
a_l-s_l & = i(-1)^l(\omega \sqrt{q})^{2l+1}Lc_l ,}}
where $L=(2l+1)!!(2l-1)!!$.  For $\omega\sqrt{q}\rightarrow 0$ this gives 
the dimensionless absorption coefficient $\sigma _l$
\eqn\absorb{ \sigma _l\equiv
{|c_l|^2\over |a_l |^2 }\simeq {4(\omega \sqrt{q})^{4l+2}\over L^2}.}

\listrefs
\end